\documentclass[12pt,fleqn,cite,epsfig]{article}

\begin{document}
\topmargin 0pt
\oddsidemargin 0mm

\renewcommand{\thefootnote}{\fnsymbol{footnote}}
\begin{titlepage}
\begin{flushright}
IP/BBSR/2002-05\\
hep-th/0205138
\end{flushright}

\vspace{5mm}
\begin{center}
{\Large \bf Intersecting Membranes from Charged Macroscopic Strings}
\vspace{6mm}

{\bf Anindya Biswas\footnote{e-mail: anindyab@iopb.res.in} 
and Kamal Lochan Panigrahi\footnote{e-mail: kamal@iopb.res.in}}\\
\vspace{5mm}
{\em Institute of Physics\\ 
Bhubaneswar 751 005, India}\\
\vspace{3mm}

\end{center}
\vspace{5mm}
\centerline{{\bf{Abstract}}}
\vspace{5mm}
We present a class of orthogonal membrane configurations
which preserve $1/4$ of the full type IIA supersymmetry. 
These membrane configurations carry additional F-string charges.
We further analyze the $D1-D3$ configuration after applying T- duality 
along the world volume directions of the above orthogonal membranes.

\end{titlepage}
\newpage
Various configurations of D-branes and their bound
states \cite{witten,li,dougla,kamal1} have emerged as important objects 
in string theory. They are very useful in understanding 
string theory and gauge theory beyond their perturbative regime. 
They are also important in understanding black holes from 
a microscopic point of view \cite{maldacena}. It is well known 
by now that a system of multiple D- branes related by $SU(N)$ 
rotation, preserves unbroken supersymmetry \cite{douglas} and, 
consequently, it is interesting to study their properties and applications
in order to test various conjectures in string theory. Explicit 
supergravity solutions for these branes and their bound states 
are studied at length in literature \cite{myers,gaunt}. 
A special class of such 
branes are orthogonally intersecting branes. The purpose of this paper
is to construct a class of orthogonally intersecting supersymmetric
branes in ten dimensional type IIA/IIB string theory. We generalize these 
configurations starting from charged macroscopic string solution. 

Charged macroscopic strings have been useful in the past in establishing
various duality conjectures in string theory. Like the neutral
solutions, BPS nature of these configurations, allow us to construct stable
multi-string configurations. Charged string solutions are in general 
constructed from neutral strings by means of a solution generating
technique. In particular, the solutions presented in \cite{sen},
are parametrized by a group $O(1,1; 1,1)$. In this case, the complete solution 
is represented by two nontrivial parameters $\alpha$ and $\beta$.
The supersymmetry of the solutions while embedded in type II string 
theories, is explicitly shown in \cite{kumar} for particular values of
$\alpha$ and $\beta$.
In general, these strings carry vectorial charges and currents
in addition to the string charges. The bound states constructed 
from these strings are interpreted as D- brane bound sates with
F- string charges being dissolved in different orthogonal 
directions \cite{kamal}.  
Starting with a particular class of D-branes of \cite{kamal}, we construct 
orthogonal membrane configurations in ten dimensions. They are
different from the ones constructed in \cite{myers1,hambli}, 
as our solutions contain NS-NS
2-form potentials, indicating the presence of F-string charges as well. 
Indeed in certain limits, our solution reduces to
\cite{myers1,hambli}. It preserves $1/4$ of the full IIA supersymmetry.
We then apply $T$- duality along one of the world volume directions to
generate bound state configuration of orthogonal (D1, D3) branes.
The existence and stability of $(D1 \bot D3)$ is checked by
examining the mass-charge relationship explicitly.
We start by writing down the supergravity solution of a D2-brane
constructed from the charged macroscopic string\cite{kamal}. 
This membrane carries NS-NS two form charge along with the R-R 
three form potential. 
These NS-NS charges are parameterized by the 
solution generating parameter $\alpha$. One can obtain the 
charge neutral D2-brane solution\cite{kastor,duff} by taking $\alpha = 0$
limit.  
\begin{eqnarray}
ds^2 &=& {(1 + X)^{3/2}\over {1 + X \cosh^2 \alpha}}\Big[{1\over {1 +
X}}\{ -{(dt)}^2 + {(dx^9)^2} + {{1 + X \cosh^2 \alpha}\over {1 + X}}
{(dx^8)}^2\} \cr
& \cr
&+& {{1 + X \cosh^2 \alpha}\over {1 + X}}
\sum_{i=1}^{7} ({dx^i})^2\Big],\cr
& \cr
{e^{\phi}}^{(10)} &=& {(1 + X)^{3/2}\over {1 + X \cosh^2 \alpha}},~~~
A^{(3)} = - {X \cosh \alpha\over{1 + X \cosh^2\alpha}}dt \wedge dx^8
\wedge dx^9,\cr
& \cr
A^{(1)} &=& - {X \sinh\alpha\over{1 + X}}dx^8,~~~
B^{(2)} = {X \sinh\alpha\cosh\alpha\over{1 + X\cosh^2\alpha}}dt \wedge dx^9,
\label{d2}
\end{eqnarray}
where $X = C ({l \over{|\vec{x}|}})^{5}$ is the harmonic function in
the transverse space and $l$ is the length parameter.
Now, we present the supergravity solution of two such membranes orthogonal 
to each other. To start with, both membranes (\ref{d2}) are lying 
in $x^8-x^9$ plane and are parallel to each other. 
A rotation between $x^8-x^9$ and $x^1-x^2$
plane is then performed following \cite{myers1}. 
The rotation angles in our case 
are $(\theta_1, \theta_2) = (0, \pi/2)$, where $\theta_1$ and
$\theta_2$ are the angles with which the two membranes are being
rotated. The second rotated brane now lies in $x^1-x^2$ plane which
is orthogonal to the $x^8-x^9$ plane. The final solution is given by:  
\begin{eqnarray}
ds^2 &=& {(1 + X)^{3/2}\over {1 + X \cosh^2 \alpha}}\Big[{1\over {1 +
    X}}\{ -{(dt)}^2 + ( 1 + X_2 \cosh^2 \alpha){(dx^9)^2}\cr
& \cr 
&+& {{1 + X \cosh^2 \alpha}\over {1 + X}}(1 + X_2){(dx^8)}^2 
+ (1 + X_1 \cosh^2 \alpha){(dx^1)}^2 \cr
& \cr
&+& {{1 + X \cosh^2 \alpha}\over {1 + X}}(1 + X_1){(dx^2)}^2\} + 
{{1 + X \cosh^2 \alpha}\over {1 + X}}\sum_{i=3}^{7} ({dx^i})^2\Big],\cr
& \cr
{e^{\phi}}^{(10)} &=& {(1 + X)^{3/2}\over {1 + X \cosh^2 \alpha}},\cr
& \cr
A^{(3)} &=& - {\cosh \alpha\over{1 + X \cosh^2\alpha}}dt \wedge
\Big[(X_1 + X_1 X_2)dx^8 \wedge dx^9 - (X_2 + X_1 X_2)dx^1 \wedge
dx^2\Big],\cr
& \cr
A^{(1)} &=& - {\sinh\alpha\over{1 + X}}\Big( X_1 dx^8 + X_2 dx ^2
\Big),\cr
& \cr
B^{(2)} &=& {\sinh\alpha\cosh\alpha\over{1 +
X\cosh^2\alpha}}dt \wedge(X_1 dx^9 - X_2 dx^1),
\label{per-D2}
\end{eqnarray}
where $X_{a} = C ({l_a \over{|\vec{x} - \vec{x_{a}}|}})^{3}$,
with $a = 1, 2$, are the harmonic functions in the tranverse space,
given by $x^i$'s, and $ X = ( X_1 + X_2 + X_1 X_2)$\cite{myers1,hambli} with
$l_a$'s being the arbitrary
positive parameters having the dimensions of length. 
To clarify further the form of $X$ appearing above, one notices that
its general structure for a system of two D2-branes, 
when their relative orientation is restricted by SU(2) rotation
is given by\cite{myers1,hambli,vijay}:
\begin{equation}
X = X_1 + X_2 + X_1 X_2 \sin^2(\theta_1 - \theta_2),
\end{equation}
with $\theta_1$ and $\theta_2$ are the angles with which the branes
being rotated. In the present context, however, 
($\theta_1, \theta_2) = (0, \pi/2)$. 
We would like to point out here, that for a configuration of
parallel membranes $X$ is the sum of individual 
harmonic functions describing each of them. But, 
when the branes are oriented at some angle, the form of $X$ gets
modified as given above. As stated earlier, this class
of solutions including NS-NS charges, are parametrized by $\alpha$. 
We have also explicitly verified that in asymptotic limit, 
the above solution indeed
satisfy the type IIA string equations of motion. Moreover, in
$\alpha=0$ limit of the above solution reduces 
to the ones given in \cite{tseyt,hambli}. To confirm further the 
correctness of our solution, we present the mass-charge relation 
(of BPS type) in the following discussion. We have therefore obtained,
an orthogonal configuration of membranes following the procedure as
described in \cite{myers}. One notices that the above solution carries 
3- form R-R potentials which are in $(x^8-x^9)$ and
$(x^1-x^2)$-plane. In addition it
also carries $F$-string charges along $x^1$ and $x^9$ directions.  
We now go on to check its supersymmetric
property by analyzing the mass-charge relationship explicitly.  

The charges associated with the membranes can be
read off from the leading behaviour of the potentials, that are
discussed earlier. To simplify the discussion further, a dimensional reduction 
along the isometry directions is performed to check the BPS properties of
our solution, as described in \cite{roy,kamal}. After dimensional 
reduction, we get charges from the following fields : 
$A^1_t \sim {\cal{B}}_{t 9}$, 
$A^2_t \sim {\cal{B}}_{t 1}$
$A^3_t \sim {\cal{A}}_{t 8 9}$ and $A^4_t \sim {\cal{A}}_{t 1 2}$,
which are as follows:
\begin{eqnarray}
Q_1 &=&  -3C\cosh\alpha\sinh\alpha~{l_1}^3,\cr
& \cr
Q_2 &=& 3C\cosh\alpha\sinh\alpha~{l_2}^3, \cr
& \cr
Q_3 &=& 3C\cosh\alpha~{l_1}^3, \cr
& \cr
Q_4 &=& -3C\cosh\alpha~{l_2}^3,
\label{charge} 
\end{eqnarray}
where, $Q_i$'s are the charges associated with $A_t^{(i)}$ with $i=
1....4$. 

The ADM mass density can be calculated by using the formula as given in
\cite{lu,myers}:
\begin{equation}
m = \int \sum_{i=1}^{9-p} n^i
\left[\sum_{j=1}^{9-p} (\partial_j h_{i j} - \partial_i h_{j j}  )
 - \sum_{a=1}^p \partial_i h_{aa} \right] r^{8 - p} d\Omega,
\label{adm-mass}
\end{equation}
where $n^i$ is a radial unit vector in the transverse space and 
$h_{\mu \nu}$ is the deformation of the Einstein-frame metric
from flat space in the asymptotic region.  
The ADM mass-density for this case is found to be:
\begin{equation}
m_{(2\bot 2)} = 3C \cosh^2 \alpha({l_1}^3 + {l_2}^3). 
\label{mass}
\end{equation}
Comparing (\ref{charge}) and (\ref{mass}), we get
\begin{equation}
m_{(2 \bot 2)}^{2} = (Q_1 - Q_2)^2 + (Q_3 - Q_4)^2.
\label{mass-charge}
\end{equation}

Now, to show this formula indeed reduces to the BPS bound
that preserves certain amount of supersymmetry, one considers 
the most general form of the Bogomol'nyi mass matrix written 
explicitly in \cite{pope}. This matrix \cite{pope}, corresponds
to the most general charges of the brane configuration. In particular,
when the charges are in 2-form tensorial representation, 
the vanishing eigen-value equation of the mass matrix as obtained in 
\cite{myers1} indicates the preservation of one-quarter supersymmetry. 
The precise form of the mass formula \cite{myers1} is given by : 
\begin{equation}
m_{\pm}^2 = (q_{i j} q_{i j} \pm {1\over 2}\epsilon_{i j k l}q_{i
  j}q_{k l}), 
\end{equation}
where $q_{i j}$'s are the nonvanishing charge densities
correspond to D- membranes.
By suitably identifying the charges in our case, with $q_{i j}$'s 
of \cite{myers1}, eqn.(\ref{mass-charge}) does imply the mass-formula
of one-quarter BPS objects. We therefore conclude that our
configuration preserves $1/4$ supersymmetry.

We now construct other interesting configurations starting from
(\ref{per-D2}). These will be constructed by using
T-duality property of string theory. In particular, we will 
consider the effect of T- duality along one of the world-volume
direction, namely along $x^9$. Following the rules, that are given in 
\cite{ortin,myers}, we get the following configurations of the
metric, dilaton, NS-NS 2-form, R-R 2-forms and 4-forms : 
\begin {eqnarray}
ds^2 &=&
{{(1+X)^{1/2}}\over(1+X{\cosh^2\alpha})}\Big[-1+{{X_1^{2}\sinh^2\alpha\cosh^2
\alpha}\over{(1+X)(1+X_2\cosh^2\alpha)}}\Big]{(dt)^2}\cr
& \cr
&+& {(1+X\cosh^2\alpha)\over{(1+X)^{1/2}(1+X_2\cosh^2\alpha)}}(dx^9)^2
+ {(1+X_2)\over(1+X)^{1/2}}{(dx^8)^2} \cr
& \cr
&+&{(1+X)^{1/2}(1+X_1\cosh^2\alpha)
\over(1+X\cosh^2\alpha)}{(dx^1)^2}+{(1+X_1)\over(1+X)^{1/2}}{(dx^2)^2}\cr
& \cr
&+&{X_1\sinh\alpha\cosh\alpha\over{(1+X)^{1/2}(1+X_2\cosh^2\alpha)}}(dt)(dx^9)
+(1+X)^{1/2}{\sum_{i=3}^7}(dx^i)^2,\cr
& \cr
{e^{\phi}}^{(10)} &=& {{1 + X} \over{1 + X_2\cosh^2\alpha}}, \cr
& \cr
A^{(4)}_{t 1 8 9} &=&
{X_1 X_2\sinh^2\alpha\cosh\alpha\over2(1+X)(1+X\cosh^2\alpha)},\cr
& \cr
A^{(4)}_{t 1 2 9}&=&
-\Big[{{X_2^{2}\sinh^2\alpha\cosh\alpha\over{2(1+X)(1+X\cosh^2\alpha)}}
- {X_2(1+X_1)\cosh\alpha\over{(1+X\cosh^2\alpha)}}}\Big],\cr
& \cr
A^{(2)}_{t 8} &=& - {{X_1(1+X_2)\cosh\alpha}\over(1+X\cosh\alpha)} , ~~~
B^{(2)}_{t 1} =-{X_2\sinh\alpha\cosh\alpha\over{(1+X\cosh^2\alpha)}},\cr
& \cr
A^{(2)}_{9 8}&=& - {X_1\sinh\alpha\over(1+X)},~~~~~~ A^{(2)}_{9 2}
=-{X_2\sinh\alpha\over(1+X)}. 
\end{eqnarray}
Looking at the solutions, one notices that the resulting
configurations is a (D1, D3)-bound state where D1 and D3 are
orthogonal to each other. Apart form the R-R charges, they also
carry F-string charges along a plane orthogonal to the D-string.
Moreover, setting $X_1 = 0$, one can verify that the solution
presented above reduces to the D3-brane solution given in
\cite{kamal}.  
  
Now we calculate the mass-charge relation for this configuration as well. 
Once again, when dimensionally reduce along all the 
isometry directions, charges arise from the following fields,
$A^1_t \sim {\cal{A}}_{t 8}$, 
$A^2_t \sim {\cal{B}}_{t 1}$,
$A^3_t \sim {\cal{A}}_{t 1 2 9}$, 
$A^4_t \sim {\cal{A}}_{t 1 8 9}$ and $A^5_t \sim g_{t 9}/g_{9 9}$.
The non-zero charges corresponding to the various field strengths and
the metric components are:
\begin{eqnarray}
Q_1 &=& 3C~l_1^3 \cosh\alpha, \cr
& \cr
Q_2 &=& 3C~l_2^3 \sinh\alpha\cosh\alpha, \cr
& \cr
Q_3 &=& -3C~l_2^3 \cosh\alpha, \cr
& \cr
P &=& -3C~l_1^3 \sinh \alpha\cosh\alpha,
\label{charge-d1-d3}
\end{eqnarray}
where $Q_i$'s and $P$ are the charges corresponding to the gauge fields
$A^i_t$,(i= 1, 2, 3) and $A^5_i$ respectively.
$P$ is the momentum in the $x^9$ direction of the 10- dimensional
theory.
The mass density of $(D1\bot D3)$ is given by:
\begin{equation}
m_{(3\bot 1)} = 3C ({l_1}^3 + {l_2}^3)\cosh^2\alpha.
\label{d1-d3-mass}
\end{equation}
The mass formula following from (\ref{charge-d1-d3}) and
(\ref{d1-d3-mass}) can be written as:
\begin{equation}
{m_{(3\bot 1)}}^2 = (Q_1 - Q_3)^2 + (Q_2 - P)^2.
\label{mass-charge2}
\end{equation}
On the basis of our earlier observations, the above formula 
again indicates that a bound of $1/4$ BPS objects is saturated.

We have therefore constructed, in this paper, a general class 
of intersecting D- brane configuration starting from charged 
macroscopic strings. Preservation of one-quarter supersymmetry 
of these orthogonal brane
configurations is shown by analyzing the BPS mass-bound explicitly.
It will certainly be interesting to show the existence and
supersymmetry properties of these bound states, when one of 
them being rotated by an arbitrary SU(2) angle. 
We hope to report on it in the future.

{\bf Acknowledgement:} We would like to thank S. Mukherji for
suggesting us the problem, for useful discussions at each step of
the work and for a critical reading of the manuscript. We are grateful
to A. Kumar for several crucial discussions, for clarifying important
points and for a critical reading of the manuscript.

\end{document}